\documentclass{elsart3}

\usepackage{amsmath}
\usepackage{graphicx}

\begin{document}

\begin{frontmatter}



\title{Two-body tunnel transitions in a Mn$_4$ Single-Molecule Magnet}


\author[label1]{W. Wernsdorfer}
\author[label2]{S. Bhaduri}
\author[label1]{R. Tiron}
\author[label3]{D. N. Hendrickson}
\author[label2]{G. Christou}

\address[label1]{Lab. Louis N\'eel, CNRS, BP 166, 38042 Grenoble Cedex 9, France}
\address[label2]{Dept. of Chemistry, University of Florida, Gainesville, Florida 32611-7200}
\address[label3]{Dept. of Chemistry and Biochemistry, University of California at San Diego, La Jolla, California 92093-0358}

\begin{abstract}
The one-body tunnel picture of single-molecule magnets 
(SMMs) is not always sufficient 
to explain the measured tunnel transitions. An improvement 
to the picture is proposed 
by including also two-body tunnel transitions 
such as spin-spin cross-relaxation (SSCR) which are mediated
by dipolar and weak superexchange interactions between 
molecules. A Mn$_4$ SMM is used as a model system.
At certain external fields, SSCRs lead to additional quantum
resonances which show up in hysteresis loop measurements 
as well defined steps.
\end{abstract}

\end{frontmatter}

Single-molecule magnets (SMMs) are one of the best 
systems for studying quantum tunneling of large moments. 
Since SMMs occur as assemblies in crystals, there 
is the possibility of a small electronic interaction 
of adjacent molecules. This leads to very small 
superexchange interactions that depend strongly on 
the distance and the nonmagnetic atoms in the 
exchange pathway. Until now, such an intermolecular 
exchange interaction has been assumed to be 
negligibly small. However, our recent studies 
on about 50 SMMs suggest that in most SMMs exchange 
interactions lead to a significant influence 
on the tunnel process. 
Recently, this intermolecular exchange 
interaction was used to couple antiferromagnetically 
two SMMs, each acting as a bias on its neighbor, 
resulting in quantum behavior different 
from that of individual SMMs~\cite{WW_Nature02}. 

In this contribution, we show that dipolar 
and/or exchange interactions can lead 
to collective quantum processes. 
The one-body tunnel picture of SMMs is 
therefore not always sufficient to explain 
the measured tunnel transitions. We propose 
to improve the picture by including also 
two-body tunnel transitions such as spin-spin 
cross relaxation (SSCR) which are mediated by 
dipolar and weak superexchange interactions 
between molecules~\cite{WW_PRL02}. 
We use here a different Mn$_4$ SMM to show that 
at certain external fields SSCRs lead to 
additional quantum resonances which show up 
in hysteresis loop measurements as 
well-defined steps. 

The single-crystal X-ray structure of

[Mn$_4$O$_3$Cl(O$_2$CCH$_3$)$_3$(dbm)$_3$] 
has been reported~\cite{Aubin96,Aubin98b,Andres00}. It
crystallizes in the monoclinic $P2_1/n$ 
space group with $Z = 4$.
The molecule has the trigonal pyramidal 
[Mn$_3^{\rm III}$Mn$^{\rm IV}$O$_3$Cl]$^{6+}$
core. A virtual $C_3$ 
symmetry axis runs through the Mn$^{\rm IV}$
and Cl atoms and defines the magnetic z-axis of each molecule.
The four molecules within a unit cell are canted at an angle of
8.97$^{\circ}$ with respect to one another. 
DC and AC magnetic susceptibility measurements 
indicate a well isolated $S = 9/2$ ground state~\cite{Aubin96,Aubin98b,Andres00}.

All measurements were performed using an 
array of micro-SQUIDs~\cite{WW_ACP_01}. 
The high sensitivity  allows us to study single 
crystals of SMM.

\begin{figure}
\begin{center}
\includegraphics[width=.4\textwidth]{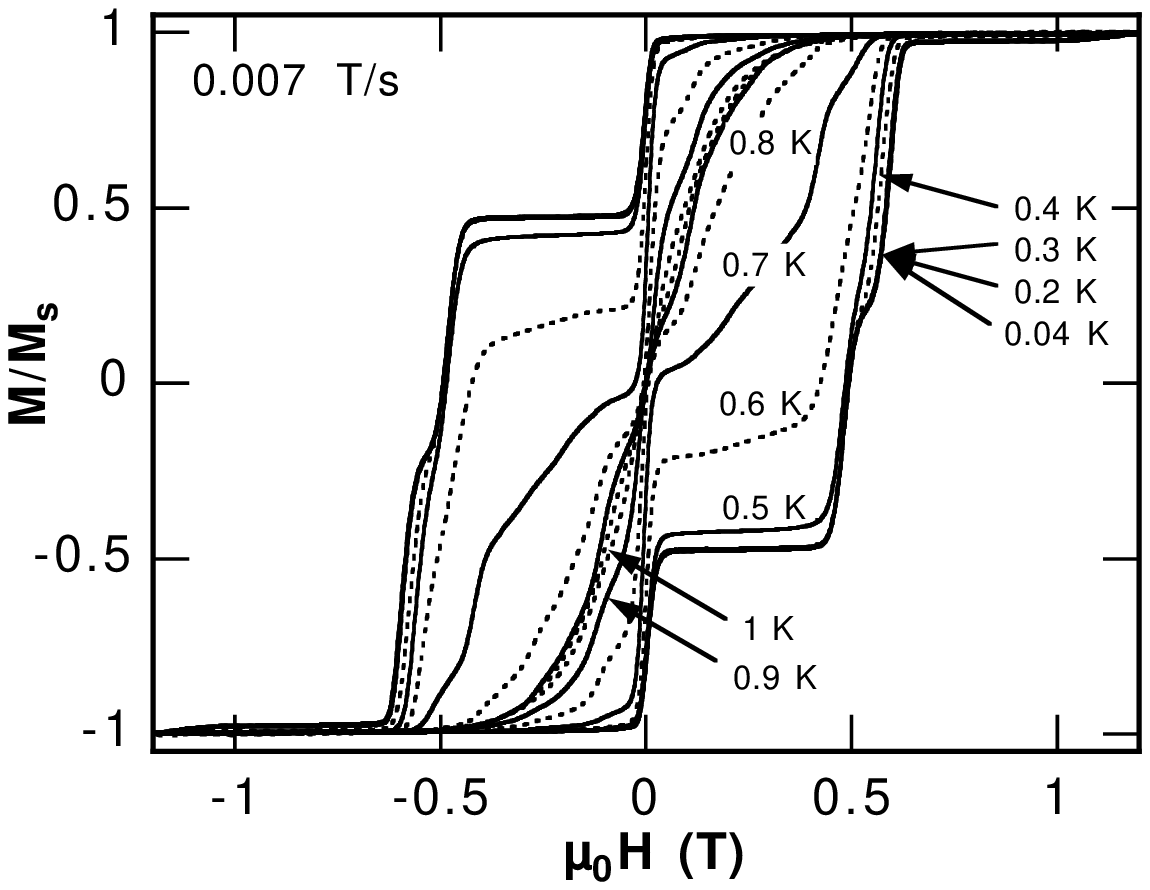}
\includegraphics[width=.4\textwidth]{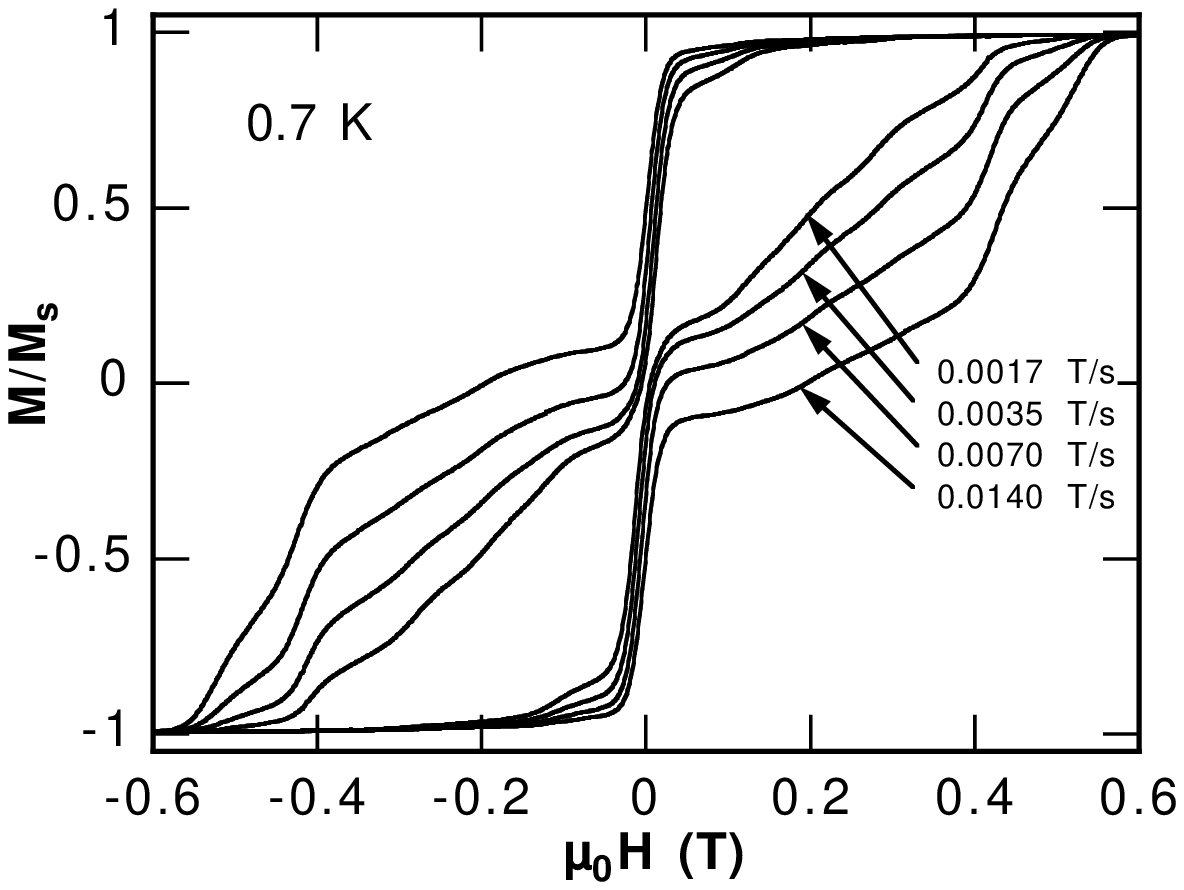}
\caption{
Normalized magnetization versus applied magnetic field. 
The resulting hysteresis loops are shown at (a) different temperatures 
and (b) different field sweep rates. 
Note that the loops become temperature-independent 
below about 0.3 K but are still sweep-rate-dependent 
owing to resonant quantum tunnelling between discrete 
energy states of the Mn$_4$ SMM.}
\label{hyst}
\end{center}
\end{figure}

Fig. 1 shows typical hysteresis loops 
for a single crystal of Mn$_4$. 
When the applied field is near an avoided level 
crossing, the magnetization relaxes faster, 
yielding steps separated by plateaus. 
A closer examination of the tunnel transitions 
however shows fine structures which cannot 
be explained by the one-body tunnel picture
(giant spin model). We suggest 
that these additional steps are due 
to a collective quantum process, called spin-spin cross-relaxation (SSCR), 
involving pairs of SMMs which are coupled by dipolar 
and/or exchange interactions. 
We used different techniques to show that 
different species due to loss of solvent or
other defects are not the reason of the observed additional 
resonance transitions.
Such SSCR processes were recently 
observed in the thermally activated regime of a LiYF$_4$ 
single crystal doped with Ho ions~\cite{Giraud01}
and on other Mn$_4$ SMMs~\cite{WW_PRL02}.

It is important to note that 
in reality a SMM is coupled to many 
other SMMs which in turn are coupled to many other 
SMMs. This represents a complicated many-body problem
leading to quantum processes involving more than two SMMs. 
However, the more SMMs that are involved, the lower 
is the probability for occurrence. 
In the limit of small exchange couplings and transverse
terms, we therefore consider only processes 
involving one or two SMMs. 
The mutual couplings between all SMMs 
should lead mainly to broadenings and small 
shifts of the observed quantum steps.

The question arises 
whether such transitions also play a role in 
other SMMs like Fe$_8$ and Mn$_{12}$. 
A diagonalization of the spin-Hamiltonian 
of such molecules shows clearly that 
SSCR should occur also. 
However, it turns out that these transitions 
are very close to the single spin tunnel 
transitions and only broaden them.
Nevertheless, such transitions should be
included in a quantitative description
of the relaxation rates, in particular
in the thermally activated regime
or for high applied fields.


\end{document}